\documentclass[11.5pt]{article}
\pdfoutput=1
\usepackage{graphicx}
\usepackage[super,sort&compress,comma]{natbib}
\usepackage[version=3]{mhchem}
\usepackage{times,mathptmx}
\usepackage{sectsty}
\usepackage{lastpage}
\usepackage[format=plain,justification=raggedright,singlelinecheck=false,font=small,labelfont=bf,labelsep=space]{caption}

\begin{document}

\title{A High Power Density, High Efficiency Hydrogen-Chlorine Regenerative Fuel Cell with a Low Precious Metal Content Catalyst}
\author{Brian Huskinson,$^{1}$ Jason Rugolo, $^{1}$ Sujit K. Mondal,$^{1}$ and
Michael J. Aziz$^{\ast}$$^{1}$}

\footnotetext[1]{Harvard School of Engineering and Applied Sciences, 29 Oxford Street, Cambridge, MA, USA. Fax: +1 617 495 9837; Tel: +1 617 495 9884; E-mail: aziz@seas.harvard.edu}

\maketitle

\begin{abstract}
We report the performance of a hydrogen-chlorine electrochemical cell with a chlorine electrode employing a low precious metal content alloy oxide electrocatalyst for the chlorine electrode: 
\ce{(Ru_{0.09}Co_{0.91})_3O_4}. The cell employs a commercial hydrogen fuel cell electrode and transports protons through a Nafion membrane in both galvanic and electrolytic mode. The peak galvanic power density exceeds 1 W cm$^{-2}$, which is twice previous literature values. The precious metal loading of the chlorine electrode is below 0.15 mg Ru cm$^{-2}$.
Virtually no activation losses are observed, allowing the cell to run at nearly 0.4 W cm$^{-2}$ at 90\% voltage efficiency. We report the effects of fluid pressure, electrolyte acid concentration, 
and hydrogen-side humidification on overall cell performance and efficiency. A comparison of our results to the model of Rugolo \emph{et al.} [Rugolo \textit{et al., J. Electrochem. Soc.}, 2012, \textbf{159}, B133] points out directions for further performance enhancement. 
The performance reported here gives these devices promise for applications in carbon sequestration and grid-scale electrical energy storage.
\end{abstract}

\section{Introduction}
\begin{figure}[t]
  \centering
  \includegraphics[height=2.7in]{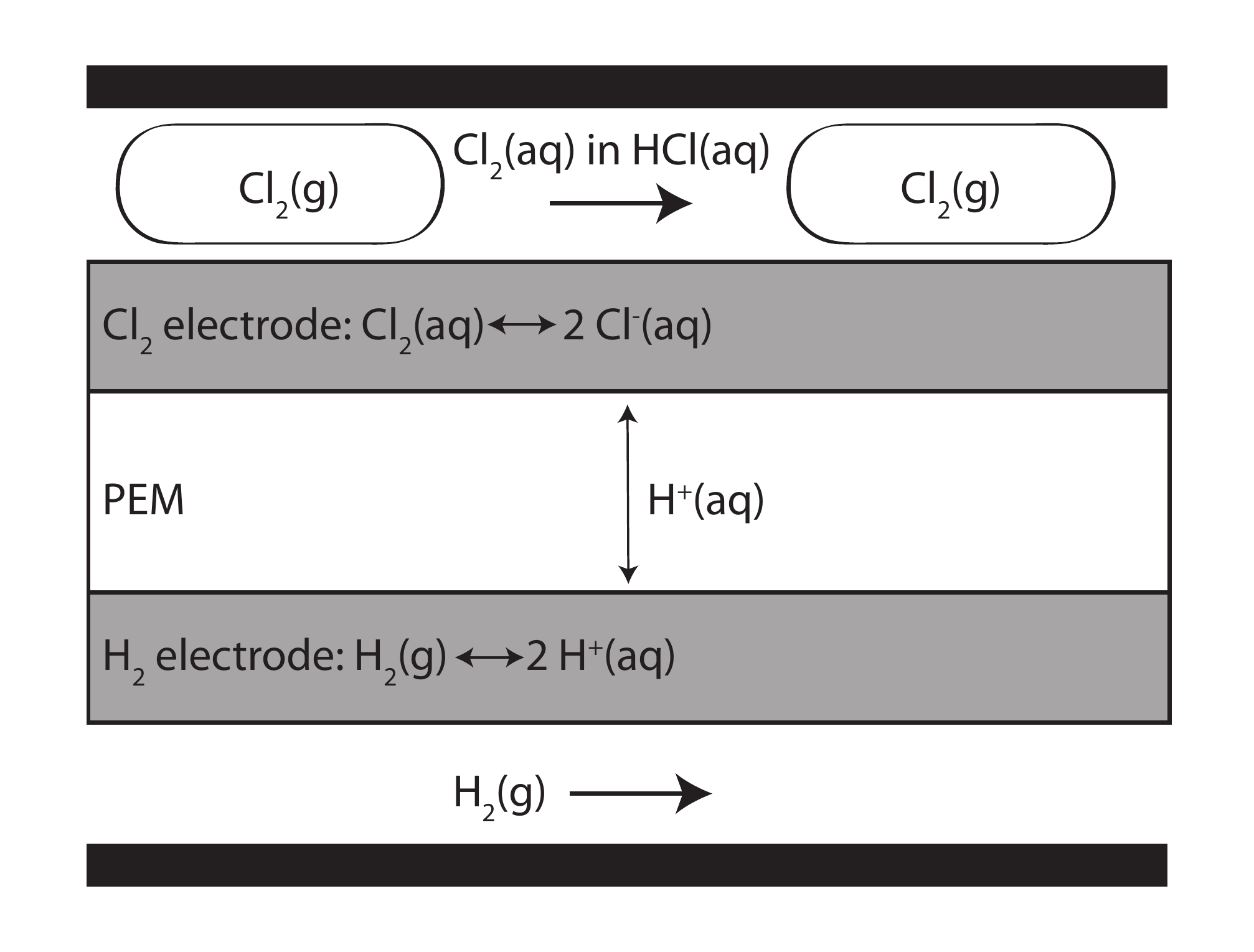}
  \caption{Schematic of a regenerative hydrogen-chlorine fuel cell, showing a cross-section along the length of a channel. In charge mode, hydrochloric acid, HCl, is electrolyzed to form \ce{Cl2} on the anode side of the cell and \ce{H2} on the cathode side. This process is not spontaneous, so a potential must be applied across the electrodes. In discharge mode, \ce{H2} and \ce{Cl2} are fed into the anode and cathode sides of the cell, respectively, spontaneously forming HCl and generating an external current. The proton exchange membrane (typically Nafion) must conduct \ce{H^+} ions in both modes (but the direction of conduction switches). If all reactants and products are stored in external tanks, forming a closed system, this reversible fuel cell can function as a flow battery.}
  \label{fig:rHHFC_schematic}
\end{figure}

Hydrogen-halogen electrochemical devices are of interest for a number of applications. Flow batteries utilizing these chemistries could be used for grid-scale electrical energy storage. \cite{Mellentine:2011gw,Eyer:2010ti,Rugolo:2012kj,Gileadi:1977un}
Cells that utilize either chlorine (\ce{Cl2}) or bromine (\ce{Br2}) have generated particular interest.\cite{Savinell:1988tn,Yeo:1980vr,Yeo:1980wr,Thomassen:2006do,Chin:1979uk,Livshits:2006} Figure \ref{fig:rHHFC_schematic} is a schematic showing how a cell operating on \ce{Cl2} would operate in both electrolytic (charge) and galvanic (discharge) modes.
\footnote[2]{We use the term ``regenerative fuel cell'' to refer to an electrochemical cell designed for operation in both galvanic and electrolytic modes in steady state,
in which the reactants and products are fluids and their activities are time-independent. If one were to take a regenerative fuel cell and add external storage tanks to hold the reactants and products,
thereby forming a closed system to be charged and discharged, then even when the cell is operated at constant current, the reactant and product activities become functions of time, or state of charge.
We refer to such a device as a ``flow battery.'' Our own device is capable of being operated in either of these manners: in the results reported here the gas pressures are constant but the hydrochloric acid concentration is not, and we do not collect the products and recycle them. Although it has some of the characteristics of each, it has more of the characteristics of a ``regenerative fuel cell.'' }
In principle, a cell involving \ce{Br2} would look the same except the chlorine gas bubbles are replaced by liquid bromine droplets. In addition to grid-scale electricity storage, the hydrogen-chlorine galvanic cell is the most immature component of an electrochemically accelerated chemical weathering process that could transfer \ce{CO2} from the atmosphere to the ocean without ocean acidification.\cite{House:2009dm,House:2007kw}

Because of relatively facile chlorine redox kinetics, \ce{H2}-\ce{Cl2} cells are capable of operation with little activation loss associated with the chlorine electrode, in marked contrast to the hydrogen-oxygen fuel cell, in which sluggish oxygen reduction kinetics lead to a substantial activation overpotential. For comparison, measured values for the exchange current density for \ce{Cl2} reduction on both smooth Pt and \ce{RuO2} are 10 mA cm$^{-2}$ and 0.01 mA cm$^{-2}$,\cite{Thomassen:2005uc} respectively,
whereas a comparable figure for \ce{O2} reduction on Pt is 10$^{-6}$ mA cm$^{-2}$. \cite{Gasteiger:2004vw}
\footnote[3]{It is worth noting that, in the presence of chlorine, Pt can dissolve by forming hydroplatinic acid (\ce{H2PtCl6}), so Pt is not a suitable material for use as a chlorine electrode material. \cite{Thomassen:2003wx}}
Exchange current densities for hydrogen oxidation/evolution reactions can be as high as 600 mA cm$^{-2}$ on structured Pt electrodes, \cite{Neyerlin:2007ww} so no significant activation losses should be expected at the hydrogen electrode.

Fast reaction kinetics at both electrodes, along with other advantages such as rapid mass transport, means that hydrogen-chlorine cells have the potential to operate at high current densities with high electric-to-electric efficiencies. Work on hydrogen-chlorine fuel cells began in the 1970s, focusing on applications in both grid-scale energy storage \cite{Yeo:1980wr,Yeo:1979wv,Chin:1979uk,Nuttall:1977ta} and in high-power applications for select missile and space programs. \cite{Balko:1981tf,Anderson:1994wn} Power densities exceeding 0.3 W cm$^{-2}$ were achieved in the late 1970s, with round-trip efficiencies at 300 mA cm$^{-2}$ of up to 75\%. \cite{Yeo:1980wr} Subsequently, power densities approaching 0.5 W cm$^{-2}$ were reached. \cite{Thomassen:2003wx} Materials stability issues limited the practical application of earlier cells, e.g. Pt dissolution in hydrochloric acid. \cite{Thomassen:2003wx} Due to the success of dimensionally stable anodes (DSAs) in the industrial production of chlorine in the chlor-alkali process, \ce{RuO2} based compounds appear promising as electrode materials for use in a halogen electrode. 

A typical DSA composition used in the chlor-alkali process is \ce{(Ru_{0.3}Ti_{0.7})O_{2}}. \cite{Bommaraju:2001tu}
DSA oxide alloys have typically been examined only in the anodic direction (i.e. oxidizing \ce{Cl^-} to \ce{Cl2}) and at relatively high pH (typically around pH 2-4), which are the conditions around which chlor-alkali membrane cells operate. \cite{Beer:1980vf,Trasatti:1984wq,Ardizzone:1982up}
Mondal \textit{et al.} examined both the cathodic and anodic behavior of a number of different ruthenium oxide alloys in hydrochloric and hydrobromic acid near pH 0 and identified the composition near \ce{(Ru_{0.1}Co_{0.9})_3O_4} as particularly promising for chlorine redox catalysis.\cite{Mondal:2011hd} To fabricate the cell reported here, we developed a method of depositing this electrocatalyst onto the fibers of Toray carbon paper.
In this work we demonstrate the successful operation of an \ce{(Ru_{0.09}Co_{0.91})_3O_4}
alloy as a catalyst material for a regenerative hydrogen-chlorine fuel cell. We observed negligible activation losses, even with precious metal loadings as low as 0.15 mg Ru cm$^{-2}$ . We observed a maximum cell power density exceeding 1 W cm$^{-2}$, which is twice as large as values previously reported with significantly higher precious metal loadings on the \ce{Cl2}-side of the cell. Furthermore, we observed a power density of approximately 0.4 W cm$^{-2}$ at 90\% galvanic efficiency. This is an important figure of merit when considering these devices for grid-scale energy storage, where round-trip electric-to-electric efficiencies are very important. The effects of \ce{Cl2} gas pressure, electrolyte acid concentration, and hydrogen electrode humidification are also reported, along with substantive comparisons to the \ce{H2}-\ce{Cl2} fuel cell model of Rugolo \emph{et al.}

\section{Experimental Methods}

\subsection{Electrode synthesis}
The chlorine-side electrode consisted of a Toray carbon paper coated with a single-phase \ce{(Ru_{0.09}Co_{0.91})_3O_4} alloy synthesized using standard wet chemical techniques. Before coating, the following protocol was used to clean the carbon paper: (1) several rinses in DI-\ce{H2O} (18.2 M$\Omega$ ultrapure, Millipore), (2) sonication in isopropyl alcohol (IPA, VWR International) for 10 min., (3) soak in hot (80 $^\circ$C) 50\% \ce{H2SO4} (reagent grade, Sigma Aldrich) for 30 min., (4) soak in hot (80 $^\circ$C) 6 M hydrochloric acid (HCl, ACS reagent grade, Sigma Aldrich) for 30 min., and (5) several rinses in DI-\ce{H2O}. The individual pieces of carbon paper were then dried and weighed. To deposit the catalyst material, a 2 cm$^{2}$ square of cleaned carbon paper was dipped in a solution of 0.1 M \ce{RuCl3} + 1 M \ce{CoCl2} in 12.1 M HCl, dried at 90 $^\circ$C for 15 minutes, then oxidized in an air furnace at 350 $^\circ$C (with a 45 min. ramp to 350 $^\circ$C and a 60 min. hold at this temperature). This procedure was repeated twice to achieve a total Ru loading of 0.15 mg Ru cm$^{-2}$. The hydrogen-side electrode used 2 cm$^{2}$ of either a standard ELAT\textsuperscript{\textregistered} gas diffusion electrode (GDE) with a Pt loading of 0.5 mg cm$^{-2}$ (Fuelcellstore.com), or a reformate anode utilizing a finely divided platinum/ruthenium alloy on carbon black (loading of about 0.6 mg Pt-Ru cm$^{-2}$, Alfa Aesar). Little difference in performance was seen between the two.

\subsection{Electrode characterization}
Micrographs of the chlorine electrode were obtained using a scanning electron microscope (SEM, Ultra55, Zeiss). Electron-dispersive spectroscopy was done using the same equipment. X-ray diffraction was done by depositing the same ruthenium-cobalt oxide alloy material used to form the electrode onto an amorphous substrate (glass microscope slides, VWR). $\theta$-$2\theta$ scans were done from $2\theta$ = 10$^\circ$ to 80$^\circ$ using a Bruker D8 Discover diffractometer. Copper K$_\alpha$ radiation was used. Lattice parameters were calculated using methods from Cullity. \cite{Cullity:1956tk}

\subsection{Fuel cell construction and test bench characteristics}
\begin{figure}[t]
  \centering
  \includegraphics[height=4.3in]{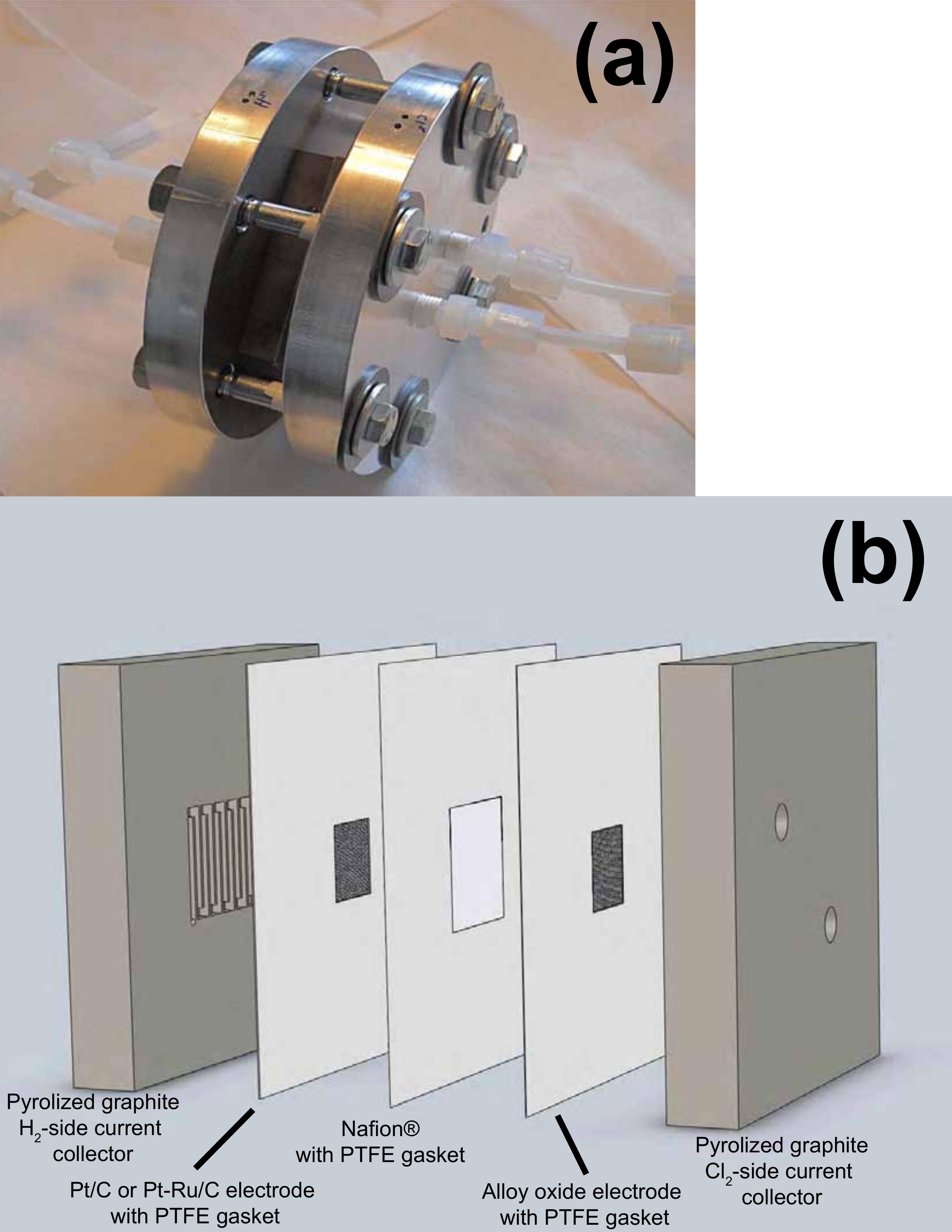}
  \caption{(a) Image of the actual cell and (b) schematic showing the internal components of the fuel cell tested. The aluminum endplates are not shown in (b).}
  \label{fig:FC_construction}
\end{figure}

The fuel cell studied here comprised a mixture of commercially available and custom-made components. Figure \ref{fig:FC_construction}a shows an image of the actual cell and  \ref{fig:FC_construction}b shows a schematic of the cell architecture. Endplates were machined out of solid aluminum. 3"x3" pyrolytic graphite blocks with single-serpentine flow channels (channel width = 0.0625 in., channel depth = 0.08 in., landing between channels = 0.031 in., Fuel Cell Technologies, Inc.) were used as current collectors. Nafion\textsuperscript{\textregistered} 112 (50 $\mu$m thick) was used as a proton-exchange membrane (PEM, Alfa Aesar), and poly-tetrafluoroethylene (PTFE) gasketing was used to seal the cell assembly. Before insertion into the cell, the following procedure was used to pretreat the PEM: (1) immersion in 85 $^\circ$C DI-\ce{H2O} for 15 min., (2) immersion in 5\% \ce{H2O2} (ACS reagent grade, Mallinckrodt Chemicals) for 30 min., (3) rinse with DI-\ce{H2O}, (4) ion-exchange twice in 0.05 M \ce{H2SO4} for 30 min. each, and (5) rinse in DI-\ce{H2O} four times, each for 15 min. Membranes were stored in DI-\ce{H2O} when not in use.

\begin{figure}[t]
  \centering
  \includegraphics[height=11cm]{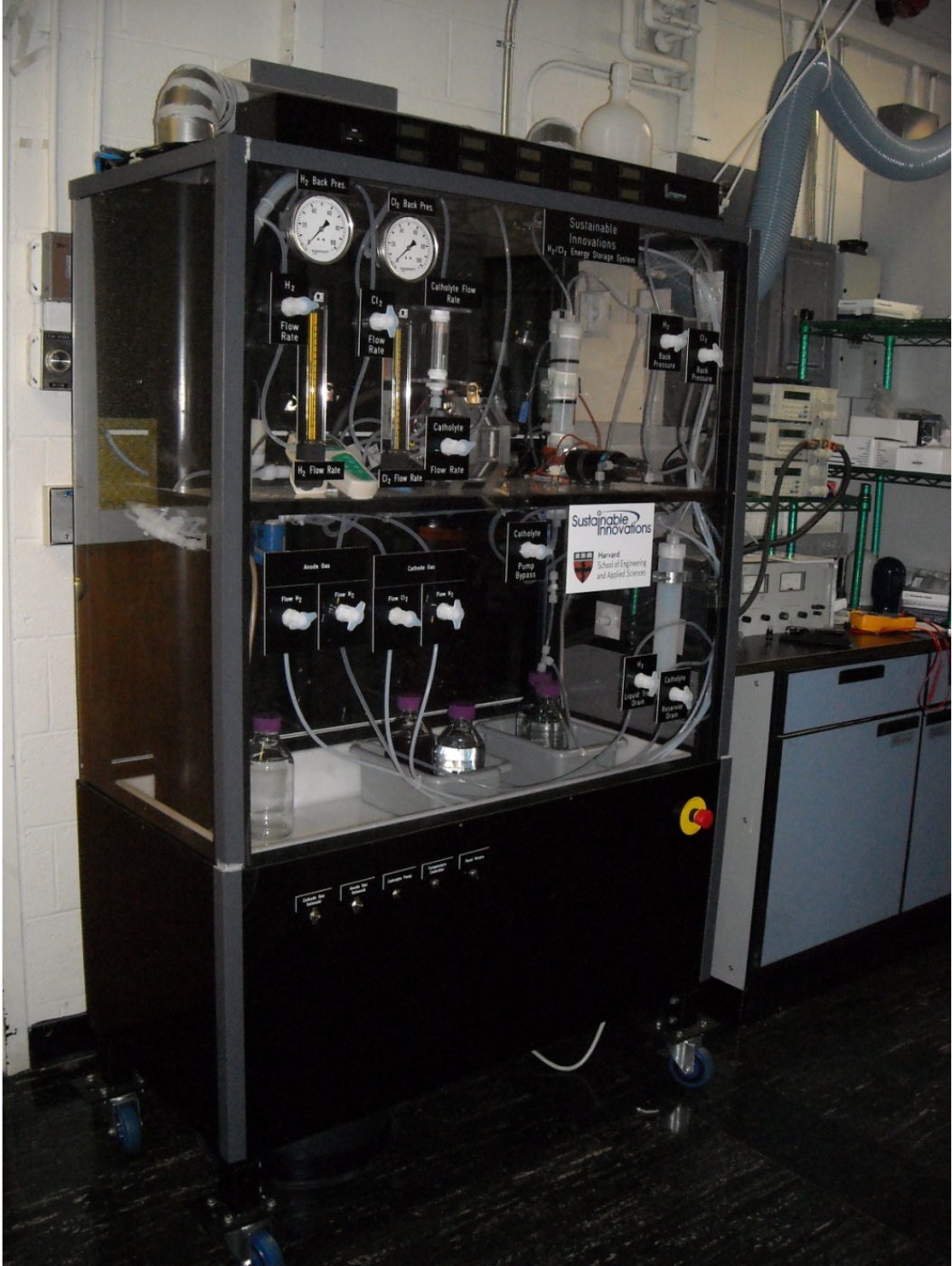}
  \caption{Image of the test bench used for cell tests. The entire cavity of the bench is held under negative pressure for safety reasons.}
  \label{fig:test_bench_photo}
\end{figure}

Six bolts (3/8''-16) torqued to 10.2 Nm completed the cell assembly, and PTFE tubing was used to transport reactants and products into and out of the cell.
Holes bored into the aluminum endplates allowed for the insertion of thermometers into each endplate to monitor the cell temperature. The cell was kept on a hotplate for temperature control. Furthermore, the liquid electrolyte reservoir was heated to improve thermal management. The system, when operating, holds about 0.8 L of electrolyte.
All measurements were conducted in a test bench designed and assembled by Sustainable Innovations, LLC. The bench exhausted to a fume hood, and all reactant gases (\ce{H2} and \ce{Cl2}) were stored inside the hood. Figure \ref{fig:test_bench_photo} shows an image of the test bench apparatus.

\ce{H2} gas (zero grade, 99.99\%, Matheson Tri-Gas) was fed to the 
hydrogen side of the cell. For non-humidified \ce{H2} electrode experiments, the dry gas stream was fed directly to the cell. In studying the effects of humidification, the dry \ce{H2} gas stream was first bubbled through a heated reservoir of DI-\ce{H2O} before entering the cell. The reservoir was maintained about 3 $^\circ$C cooler than the actual cell to prevent condensation of \ce{H2O} within the cell, which could potentially flood the hydrogen electrode.
\ce{H2} flow rates were maintained between 50-100 mL/min.

On the chlorine side, a two-phase flow was fed into the cell (see Fig. \ref{fig:rHHFC_schematic}). This consisted of a liquid phase of dilute hydrochloric acid  saturated with \ce{Cl2(aq)}, along with bubbles of \ce{Cl2} gas (high purity, 99.5\%, Matheson). Having a liquid phase meant there was continuous transport of \ce{H2O} into the cell and to the membrane surface, thereby keeping it hydrated. This allowed for the use of dry \ce{H2} gas on the hydrogen side (note that the effect of humidifying the \ce{H2} side is discussed later). The cell pressure was controlled via needle valves on the outlet tubes from the cell, and, in general, the \ce{H2} side pressure was maintained at 5 psi higher than the \ce{Cl2} side of the cell. A centrifugal pump was used to circulate HCl into and out of the \ce{Cl2} side of the cell. By teeing the pump outlet line into the \ce{Cl2} gas flow line, two-phase flow was achieved. It is worth noting that, because the electrolyte is continuously recirculated, the HCl concentration increases with time as current is being drawn from the cell. Because the electrode areas are relatively small (2 cm$^{2}$), though, the cell has to run for long periods of time before there are significant concentration changes.\footnote[4]{Running at 1 A cm$^{-2}$, for example, it would take nearly 11 hours of operation to change the electrolyte concentration by 1 M.} Acid flow rates were maintained around 100 mL/min. for all tests. As part of the results reported here, we test cell performance at three different values of the initial HCl concentration (0 M, 1 M, and 2 M).

\subsection{Fuel cell measurements}

Before introducing either reactant into the cell, a nitrogen (\ce{N2}, ultra high purity, Matheson) purge of the entire system was done, taking the system up to operating pressures and temperatures to ensure no leaks were present and to flush any \ce{O2} from the system. Reactants were then introduced on both sides of the cell, making sure that no significant pressure differential (i.e. $>$10 psi) develops across the membrane as the reactant pressures are increased. Once reactant flows, pressures, and the system temperature had stabilized, a DC electronic load (Circuit Specialists, Inc.) was used to draw current from the cell. An independent Fluke multimeter was used to make voltage readings across the endplates of the cell. For collection of voltage vs. current density curves, a given current would be drawn from the cell, and the voltage, once stabilized, would be read from the multimeter. Typically, the voltage stabilized immediately, but, as the limiting current density was approached, the voltage values became much less stable, leading to measurement difficulties. Measurements were repeated at least three times (and sometimes up to five) to reduce this error. For electrolytic operation, the cell was connected to a DC regulated power supply (Circuit Specialists, Inc.) to apply a potential across the cell. A multimeter was again used for independent voltage measurements.

\section{Results and discussion}

\subsection{Electrode characterization}
\begin{figure}[t]
  \centering
  \includegraphics[height=6cm]{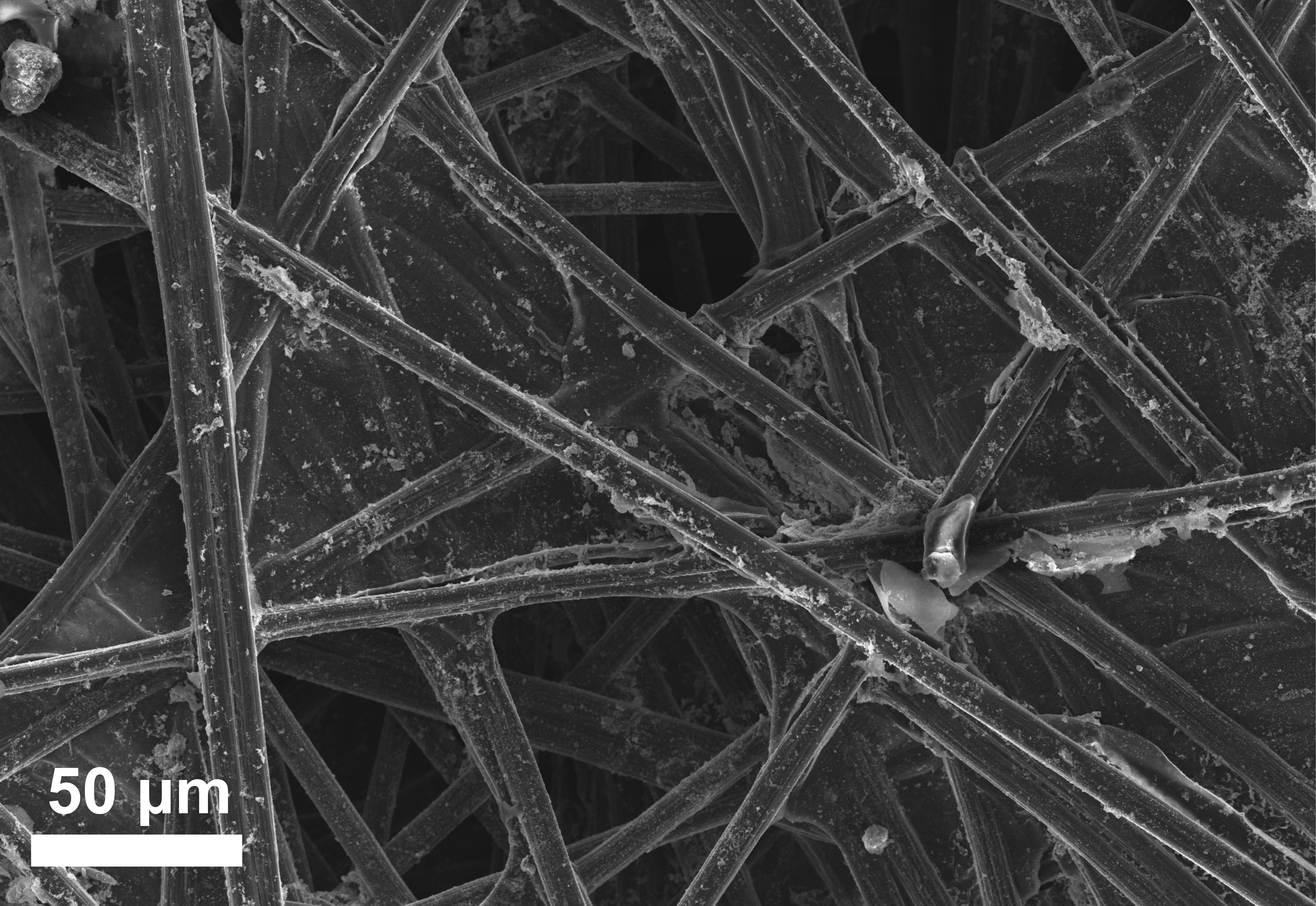}
  \caption{SEM micrograph of a ruthenium-cobalt oxide alloy electrode as deposited. An \ce{(Ru_{0.09}Co_{0.91})_3O_4} oxide alloy was deposited on a Toray carbon paper substrate, with a loading of 0.15 mg Ru cm$^{-2}$.
}
  \label{fig:SEM}
\end{figure}

A micrograph of the electrodes as deposited can be seen in Figure \ref{fig:SEM}. The \ce{(Ru_{0.09}Co_{0.91})_3O_4} oxide alloy forms highly non-uniform polycrystalline clusters on the surface of individual carbon fibers. Electrodes were made with loadings typically around 0.15 mg Ru cm$^{-2}$.

\begin{figure}[t]
  \centering
  \includegraphics[height=4.5in]{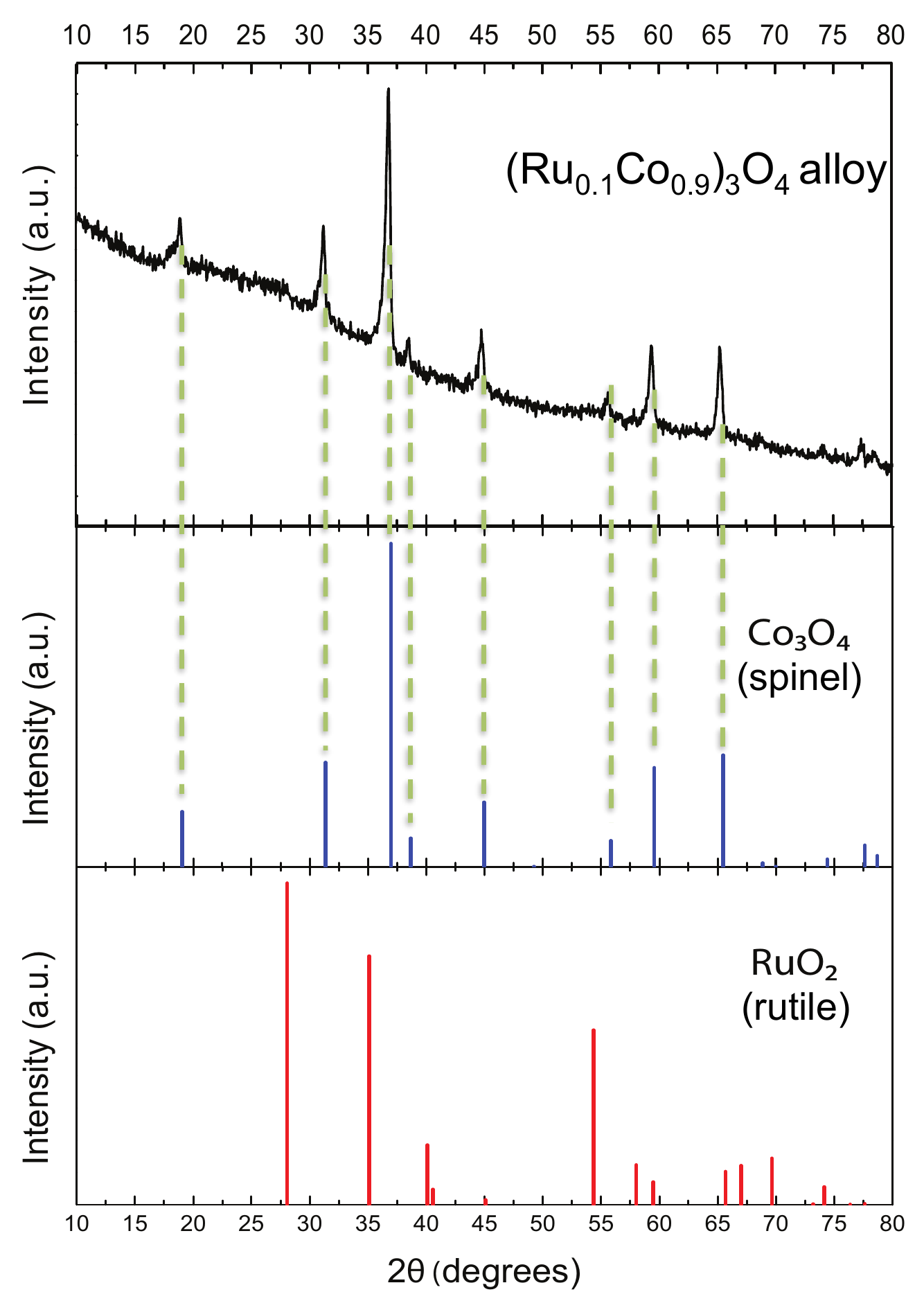}
  \caption{X-ray diffraction pattern
of an \ce{(Ru_{0.09}Co_{0.91})_3O_4} oxide alloy on a glass substrate (top), with expected patterns for pure spinel \ce{Co3O4} (middle) and rutile \ce{RuO2} (bottom) also shown. Results indicate a single-phase oxide alloy with characteristics very similar to \ce{Co3O4} (which adopts a normal spinel structure) has been created. The calculated lattice constant for the oxide alloy is 8.089 $\AA$}.
  \label{fig:XRD_RuCo_Co3O4_RuO2}
\end{figure}

XRD patterns were obtained to help determine the crystal properties. As can be seen in Fig. \ref{fig:XRD_RuCo_Co3O4_RuO2}, the wet chemical synthesis method used to form the ruthenium-cobalt oxide leads to the formation of a single-phase alloy. The alloy adopts the normal spinel structure of \ce{Co3O4}, with Ru atoms substitutionally replacing Co atoms within this crystal structure. Because Ru atoms have a larger radius than Co atoms for a given oxidation state and coordination number in a crystal, we would expect the alloy to have a larger lattice constant than that of a pure \ce{Co3O4} crystal in order to accommodate the larger Ru atoms. The calculated lattice constant for the alloy is 8.089 $\AA$, whereas that for the pure \ce{Co3O4} spinel is 8.084 $\AA$, consistent with the above reasoning.

\subsection{Cell performance}
\begin{figure}[t]
  \centering
  \includegraphics[height=6.5cm]{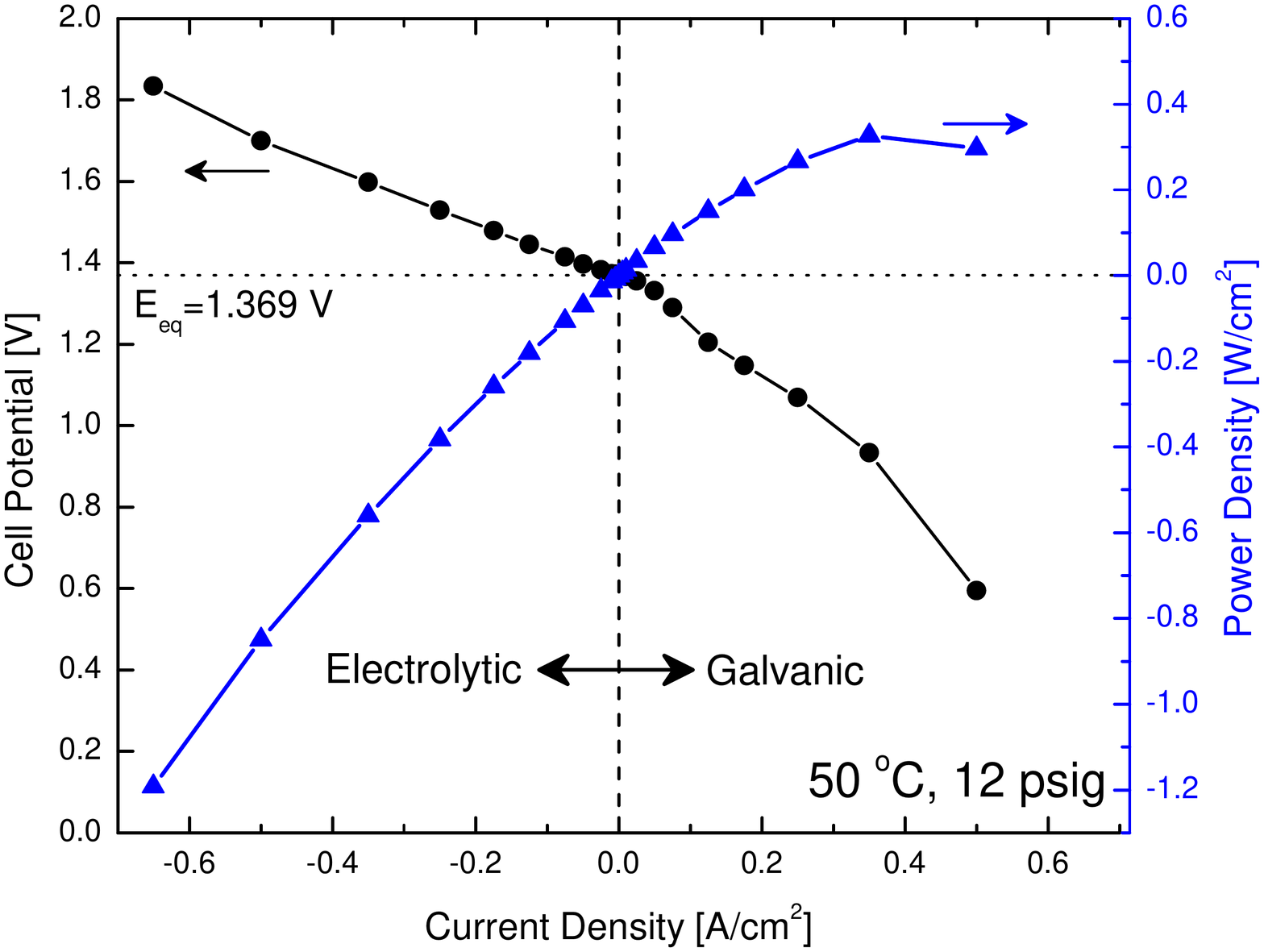}
  \caption{Potential vs. current density and power density vs. current density for the cell operated in both galvanic and electrolytic mode at 50 $^\circ$C and a \ce{Cl2} pressure of 12 psig. The cell equilibrium potential is denoted with a horizontal dotted line, and electrolytic and galvanic operation are separated by a vertical dashed line.}
  \label{fig:electrolysis-IV-PI}
\end{figure}

The performance of the cell when operated in both the galvanic and electrolytic directions at 50 $^\circ$C and 12 psig \ce{Cl2} pressure is shown in Fig. \ref{fig:electrolysis-IV-PI}. The power density is also reported. Notice that the data are smooth as the cell moves from galvanic to electrolytic operation, indicating good reversibility for the reaction and acceptable membrane performance in both modes. The increasingly rapid drop in voltage with increasing galvanic current density is an indication that we are approaching the mass transport-limited current density, at which the rate of reactant (\ce{Cl2}(aq)) diffusion through the acid bubble wall limits the reaction rate.\footnote[5]{As is common in the hydrogen-oxygen fuel cell community, we presume that \ce{H2}(g) transport is so fast as to never be the rate-limiting mass flux} In results reported subsequently, we show that increasing the chlorine gas pressure increases the limiting current density by raising the concentration of dissolved \ce{Cl2}(aq). In contrast, in electrolytic mode, with increasing current density we do not encounter a mass transport limitation. In this case the reactant is \ce{Cl-}, which is present in high concentrations.

In practice, there are two reasons why one will not want to operate at too high a voltage in electrolytic mode despite the absence of a mass-transport limitation. First, the voltage efficiency of the cell, $E_{eq}/E_{cell}$ in electrolytic mode, drops with increasing electrolytic current density. Second, the coloumbic efficiency of the electrolytic reaction will decrease with increasing cell voltage due to the propensity to evolve \ce{O2} at the chlorine electrode at higher voltages. Note that chlor-alkali cells generate 0.5-0.8 vol\% \ce{O2} in the product stream when operating around 0.4 A cm$^{-2}$,\cite{Beckmann:2001uo} so coloumbic efficiencies in this cell should remain above 99\% over a large operational range.

\begin{figure}[h!]
  \centering
  \includegraphics[height=6.5cm]{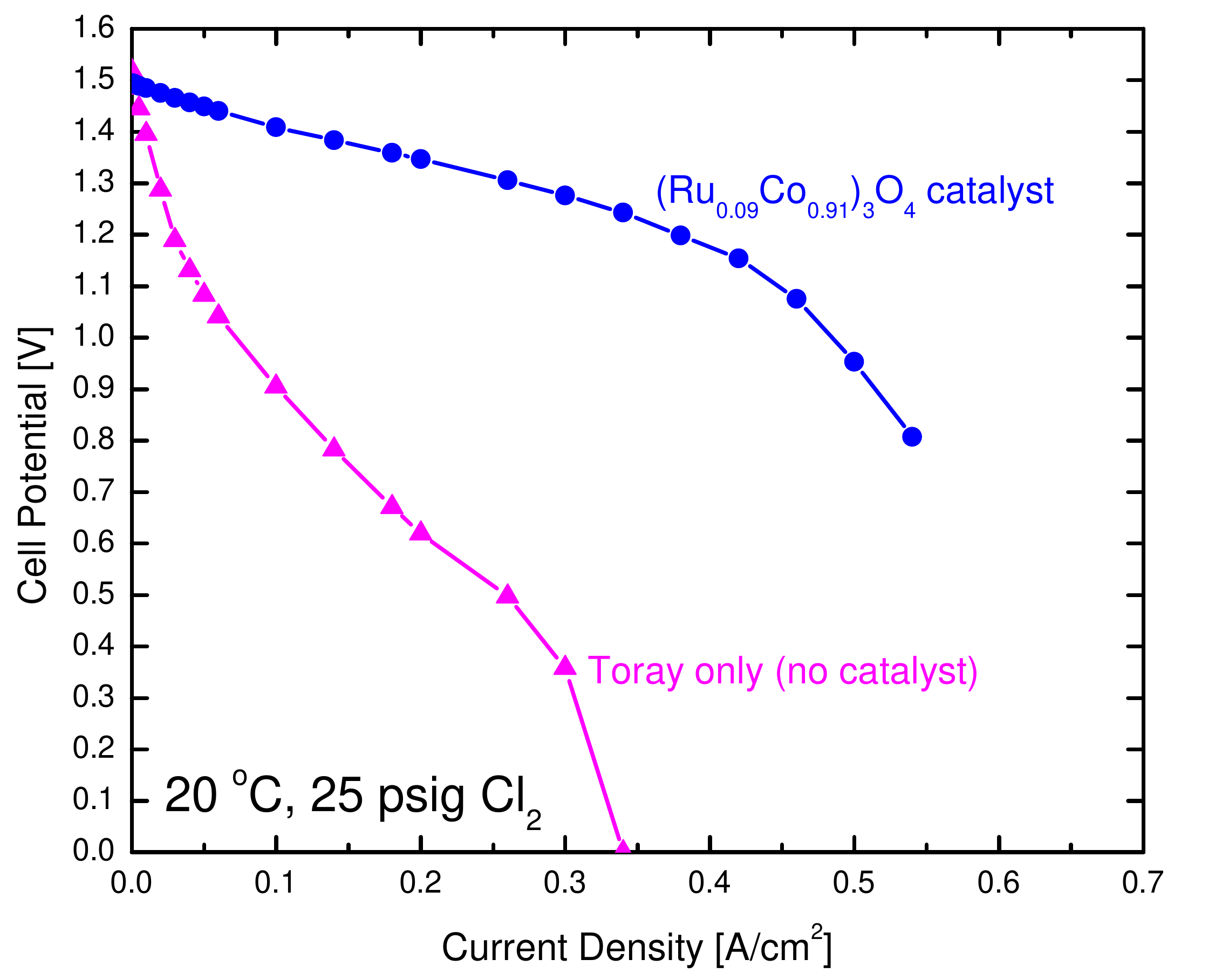}
  \caption{Cell potential vs. current density for a cell with no catalyst material on the \ce{Cl2} electrode and a cell with the ruthenium-cobalt oxide alloy described previously. An ELAT\textsuperscript{\textregistered} GDE was used on the hydrogen side in both experiments.}
  \label{fig:Toray-comparison-IV}
\end{figure}

The importance of the chlorine electrocatalyst is illustrated in Figure \ref{fig:Toray-comparison-IV}. In the absence of the catalyst material, a large overpotential is observed, which is characterized by concave-upward curvature in the potential vs. current density and is associated with sluggish charge-transfer kinetics at the chlorine electrode. In the presence of the catalyst, this overpotential is insignificant. The nonzero slope over the lower current densities in the catalyzed curve is primarily due to ohmic loss in the Nafion membrane, as we shall show later. The hydrogen-side catalyst is unchanged between the two cells: the hydrogen electrode overpotential is insignificant.

\begin{figure}[h!]
  \centering
  \includegraphics[height=5.5in]{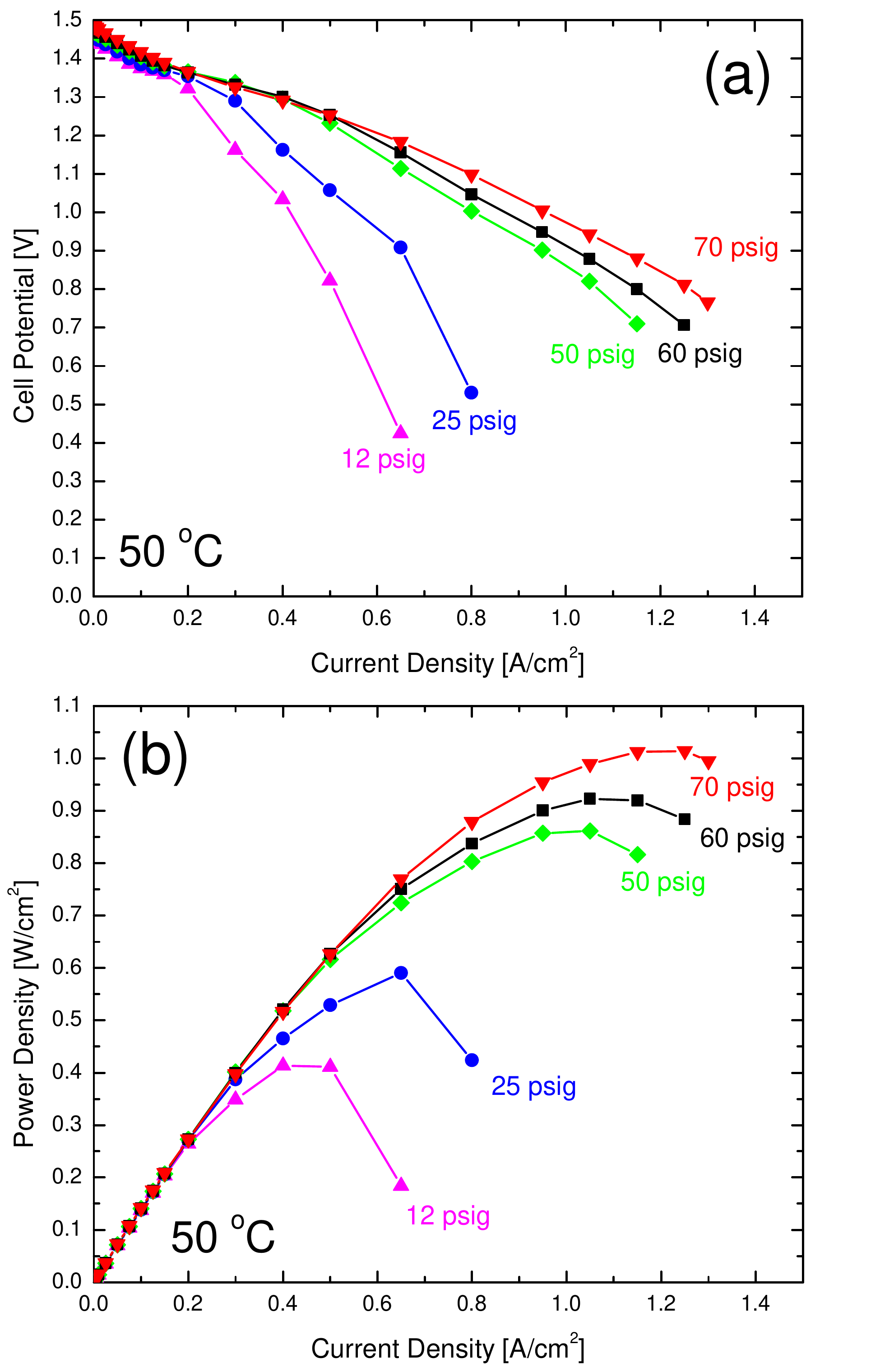}
  \caption{
(a) Cell potential vs current density, and (b) power density vs. current density for five different cell pressures. 
Data were collected at 50 $^\circ$C, and the hydrogen electrode was humidified.}
  \label{fig:humid_complete_IV_PI}
\end{figure}

Cell performance in the galvanic direction was characterized over a wide range of operating conditions. Results in Fig. \ref{fig:humid_complete_IV_PI} show the effect of changing the cell pressure. The difference is very pronounced: increasing the pressure from 12 psig to 70 psig results in a maximum power density increase from 0.41 W cm$^{-2}$ to 1.01 W cm$^{-2}$ (at 50 $^\circ$C). Pressure has almost no effect on cell performance below current densities of 0.2 A cm$^{-2}$. This is due to the fact that the primary impact of increasing cell pressure is to drive more \ce{Cl2}(g) into solution, thereby increasing its concentration and improving mass transport to the electrode-solution interface. \cite{Rugolo:2012kf} At low current densities, however, mass transport losses are insignificant, and therefore changes in cell pressure are inconsequential. Another noteworthy feature of the potential vs. current density plots in Fig. \ref{fig:humid_complete_IV_PI} is the absence of a significant activation loss associated with the electrode charge-transfer kinetics, i.e. the current-potential curves are nearly linear at low overpotentials. 
Small differences in the cell equilibrium potential can be attributed to both differences in temperature and in the activity of \ce{Cl2} as a function of pressure, in accordance with the Nernst equation.

\begin{figure}[h!]
  \centering
  \includegraphics[height=7.5cm]{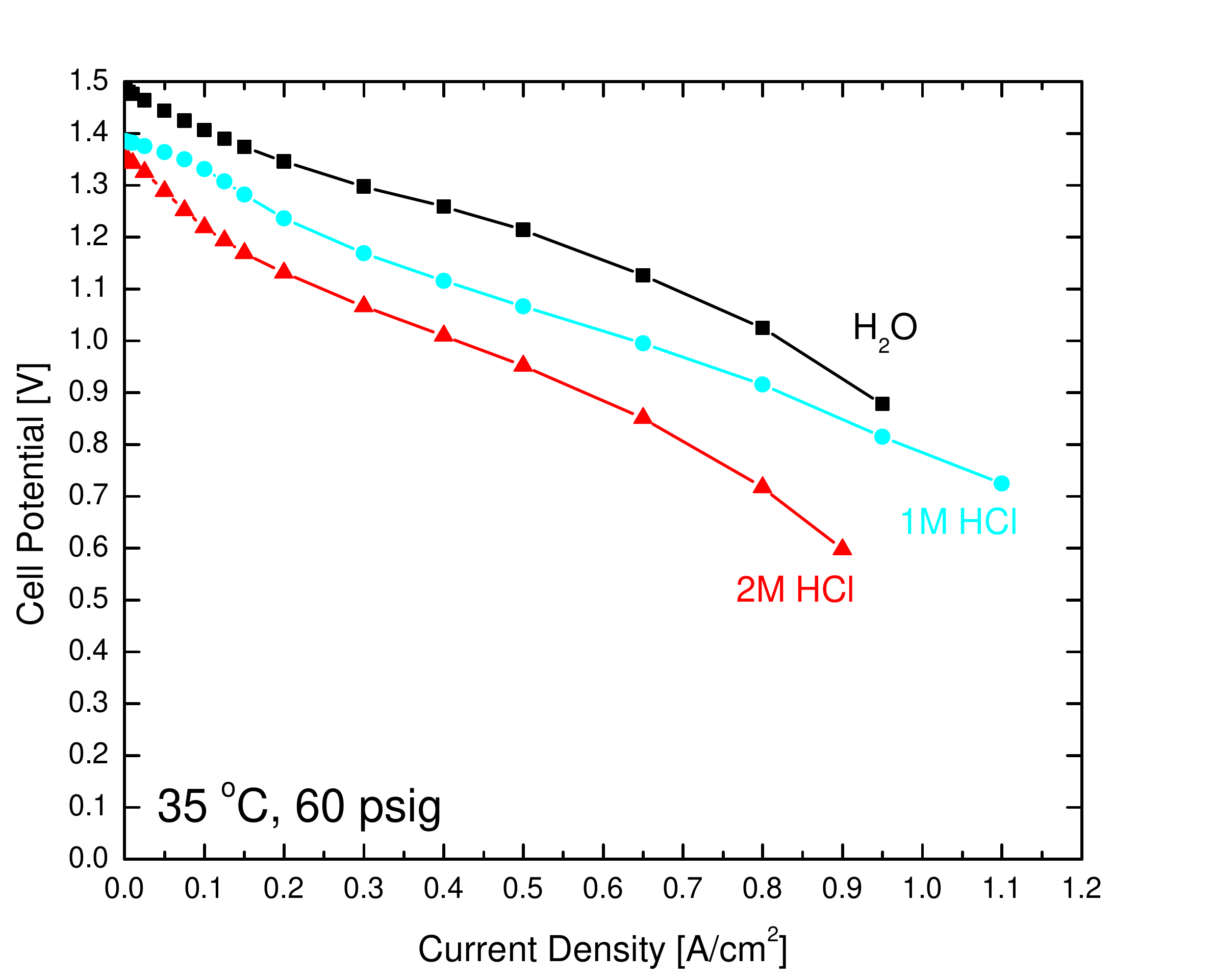}
  \caption{Cell potential vs. current density showing the effects of acid molarity on performance. The initial electrolyte concentration is indicated in the figure. “\ce{H2O}” refers to an initial electrolyte of pure water.}
  \label{fig:acid_comp}
\end{figure}

Another important characteristic of an \ce{H2}-\ce{Cl2} cell is the performance dependence on acid concentration. Acid concentration affects a number of processes occurring in the cell. First, the PEM conductivity is a function of acid concentration. This conductivity should peak at about 2.3 M HCl.\cite{Yeo:1979wv} Our results, however, indicate that the PEM conductivity in this cell is only a weak function of acid concentration: this is apparent by the approximately equal slopes of the three lines in Fig. \ref{fig:acid_comp}. 
One potential explanation for this observation is that, because HCl is being generated at the electrode-membrane interface when the cell is operated in galvanic mode, the membrane is exposed to an effective acid concentration much higher than the bulk acid concentration, regardless of the value of the bulk concentration.

\begin{figure}[h!]
  \centering
  \includegraphics[height=8cm]{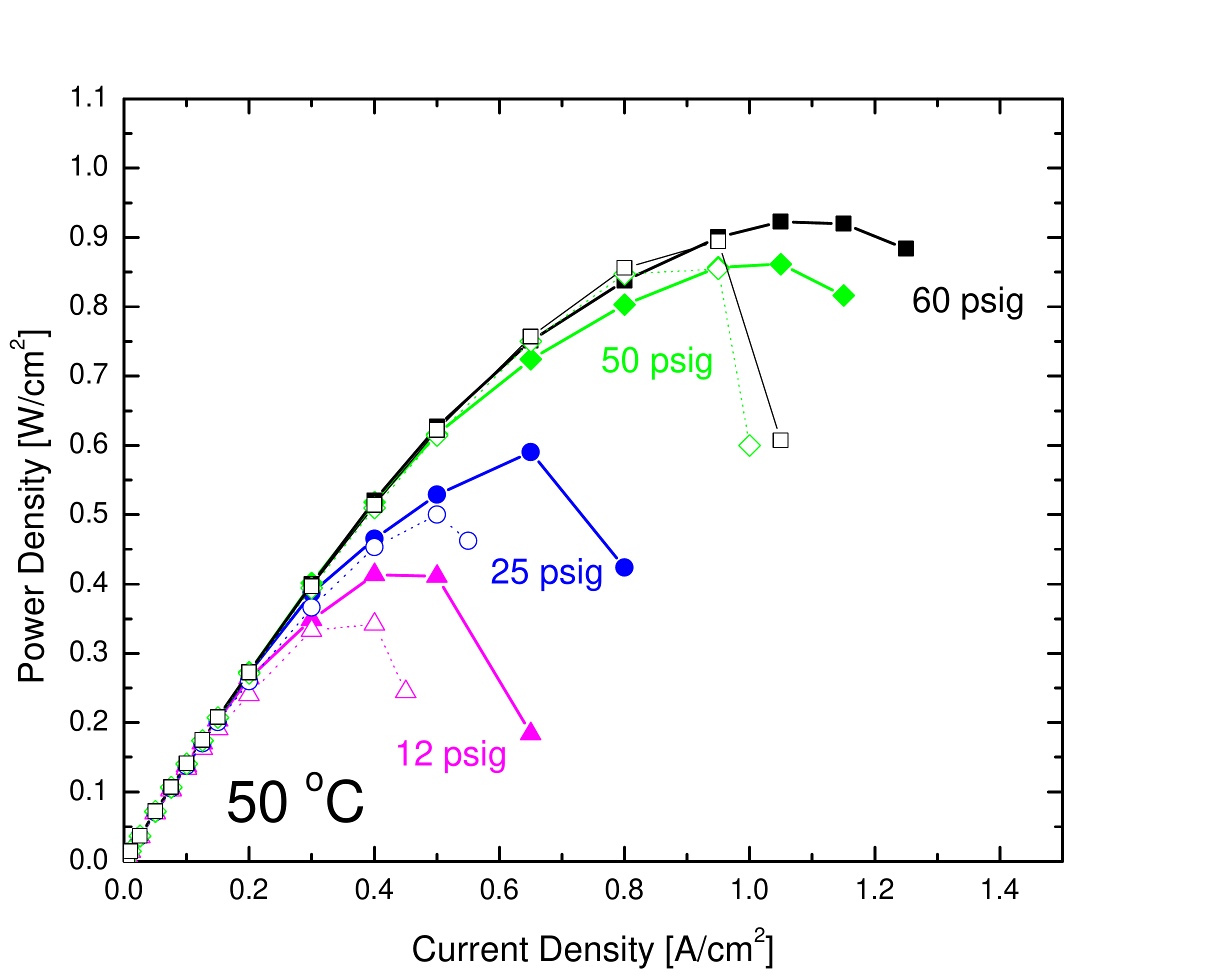}
  \caption{Effect of \ce{H2} side humidification on cell performance. Dry anode data are shown with hollow markers and dashed lines at each of the pressures studied. Humidified anode data str shown with solid markers and solid lines.}
  \label{fig:humid_comp}
\end{figure}

Hydrogen electrode humidification appears to improve cell performance, as can be seen in Fig. \ref{fig:humid_comp}. Increases in the maximum power density and the limiting current density of the cell were observed over the entire range of \ce{Cl2} pressures explored. This is likely due to membrane dehydration becoming more of an issue at high current densities. Since the membrane conductivity is a strong function of its level of hydration, drying out the membrane has the effect of increasing resistive losses through the cell.
Because there are two primary fluxes governing the water content of the membrane -- a current-independent diffusive flux from the \ce{Cl2} side (which is wet and therefore has a high \ce{H2O} activity) to the \ce{H2} side and, in galvanic mode, a current-dependent electro-osmotic flux from the \ce{H2} side to the \ce{Cl2} side -- at large current density the electro-osmotic flux may become sufficiently large to dehydrate the membrane. 
In humidifying the \ce{H2} electrode, this problem can be alleviated by delivering more water to the membrane via the \ce{H2} gas stream. This likely explains the benefits seen from \ce{H2} electrode humidification in Fig. \ref{fig:humid_comp}.

\subsection{Cell efficiency characteristics}

Because chlorine (\ce{Cl2}(aq)) crossover through Nafion is known to be minimal \cite{Yeo:1979wv} and we restrict our cell potential to a range in which oxygen evolution is slow,\cite{Beckmann:2001uo}
we expect the energy conversion efficiency of this cell to be nearly indistinguishable from the voltage efficiency, $E_{cell}(I)/E_{eq}$ in galvanic mode and $E_{eq}/E_{cell}(I)$ in electrolytic mode.

\begin{figure}[h!]
  \centering
  \includegraphics[height=8cm]{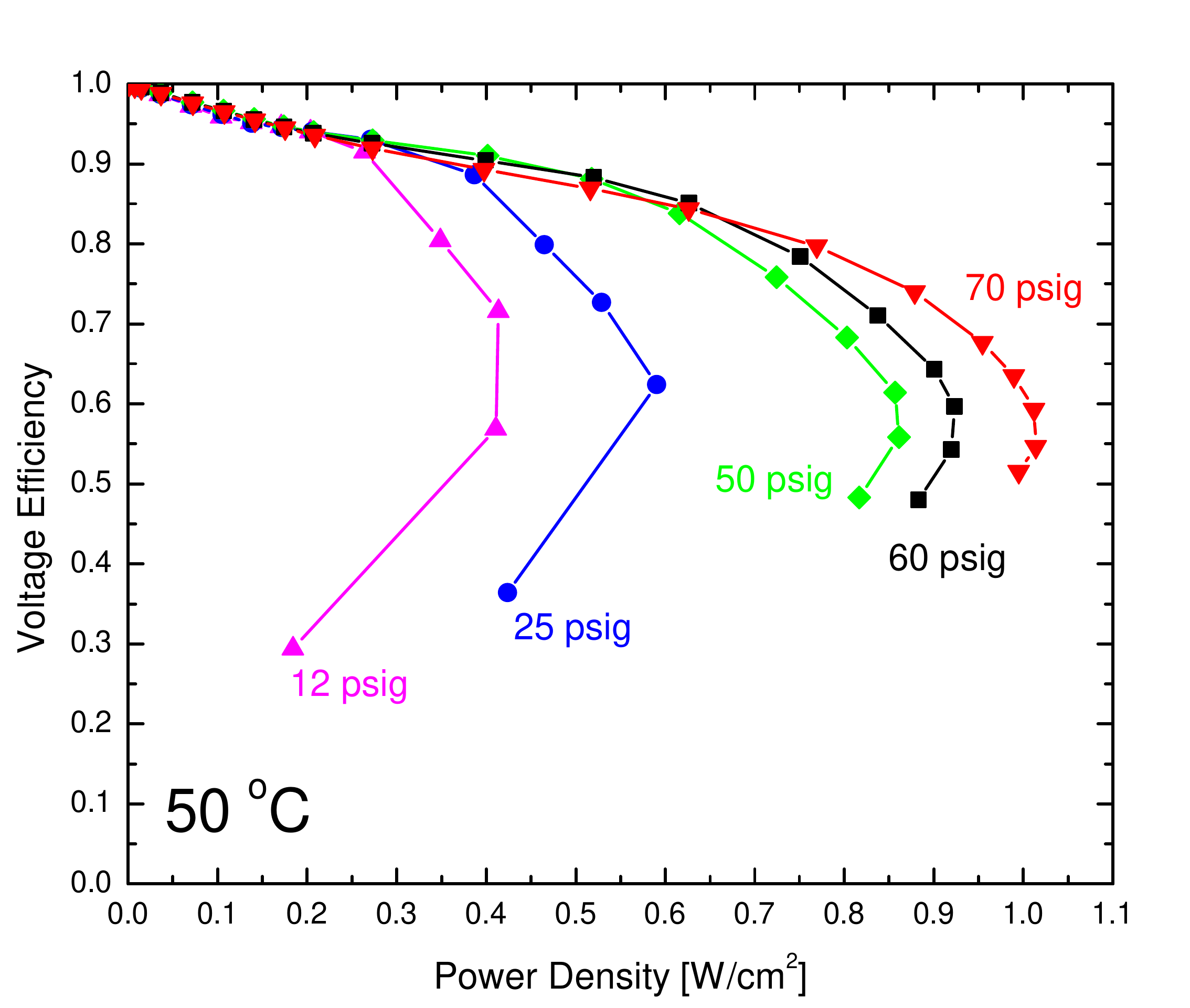}
  \caption{One-way galvanic efficiency vs. power density for five different \ce{Cl2} pressures at 50 $^\circ$C with \ce{H2} side humidification.}
  \label{fig:eff-vs-p}
\end{figure}

In Figure \ref{fig:eff-vs-p} we show the voltage efficiency as a function of power density for five different \ce{Cl2} pressures. This format is particularly useful because it illustrates the tradeoff, as one varies the operating conditions, between the two most important figures of merit of the cell. Note that a power density of approximately 0.4 W cm$^{-2}$ is reached at 90\% efficiency for all pressures exceeding about 25 psig -- the minimum pressure so that \ce{Cl2}(aq) solubility does not cause a significant mass transport limitation for this particular cell. Second, a peak power density exceeding 1 W cm$^{-2}$
was achieved at an efficiency around 56\% for 70 psig. A separate experiment reached a peak power density of 1.15 W cm$^{-2}$, but this was not part of a series so it does not appear in the figures.

\subsection{Comparisons of performance to an \ce{H2}-\ce{Cl2} model}

Comparing the experimental results to the hydrogen-chlorine fuel cell model of Rugolo \emph{et al.} \cite{Rugolo:2012kf} provides valuable insight into the behavior of the system and indicates directions for, and ultimate limitations to, further improvement in the performance of a cell of this design.
The model accounts for voltage losses due to electrocatalytic activation at both electrodes, ohmic loss in the PEM, and \ce{Cl2}(aq) mass transport through the chlorine gas bubble wall to the chlorine electrode. It predicts the voltage vs. current density behavior as one varies the following \emph{Engineering Parameters} (EPs): PEM thickness, cell pressure, bubble wall thickness, and exchange current densities at both electrodes; and the following \emph{Operating Parameters} (OPs): temperature and acid concentration.

\begin{figure}[h!]
  \centering
  \includegraphics[height=5.5in]{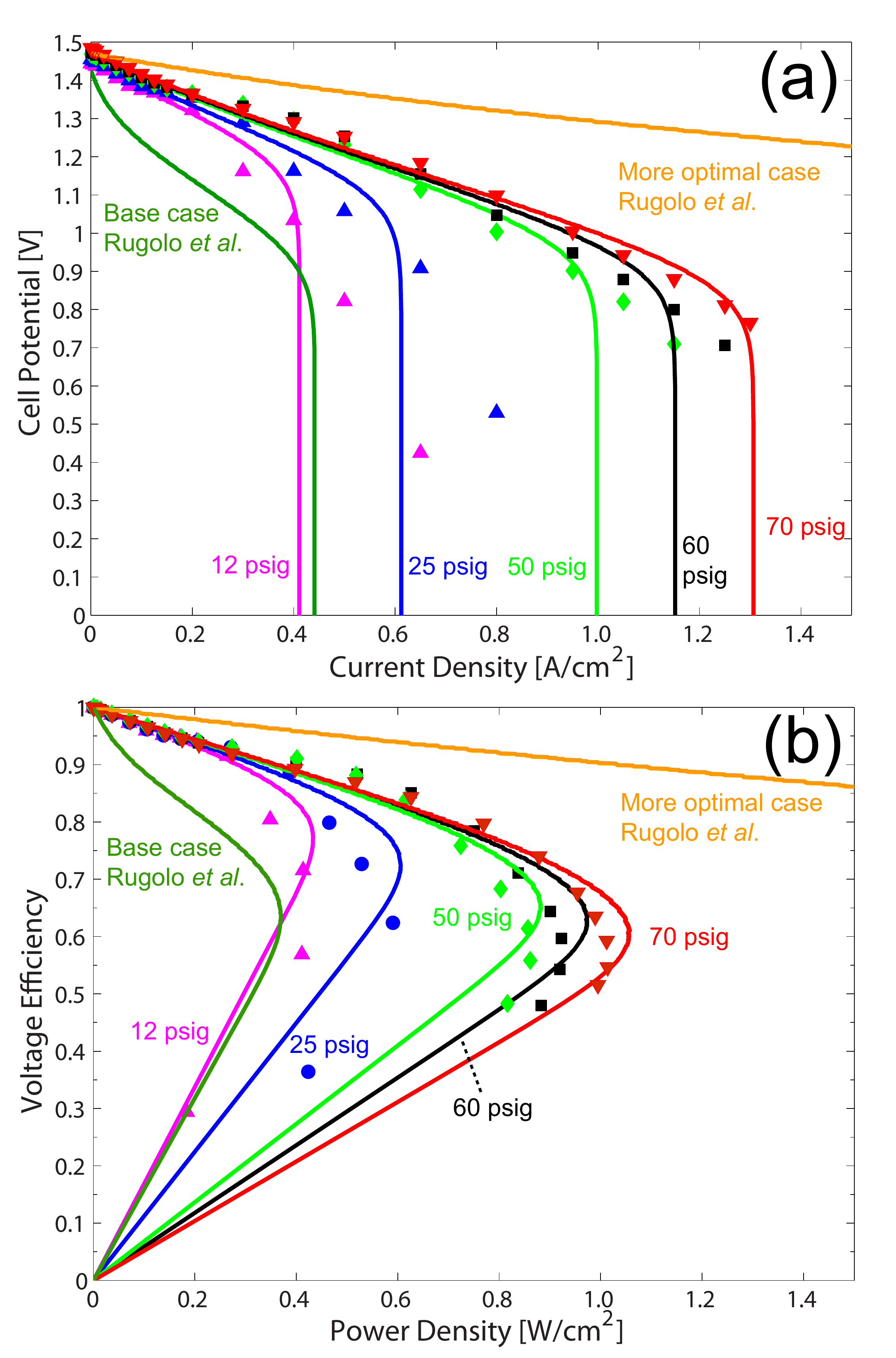}
  \caption{Comparison of model to experiment. Experimental data are indicated with symbols and model fits are shown using lines of corresponding color. (a) Cell potential vs. current density and (b) Voltage efficiency vs. power density. Also indicated are the ``Base case'' and ``More Optimal case'' scenarios described by Rugolo \emph{et al.}\cite{Rugolo:2012kf}} 
  \label{fig:comparison-to-model}
\end{figure}

We fit the model to our data using known values of all of these parameters\footnote[6]{The value for the hydrogen electrode exchange current density was held at 250 mA cm$^{-2}$, consistent with results in work by \citealt{Neyerlin:2007ww} We control directly the other known, non-adjustable parameters.} but three, which were treated as adjustable parameters. (1) The chlorine exchange current density, $i_{0}^{Cl}$, was chosen to be 175 mA cm$^{-2}$ in order to fit the data.
(2) A series resistance\footnote[7]{For dimensional consistency, the value of this resistance must be multiplied by the cell area.}, absent from the original model, accounting for ohmic losses through the current collectors, endplates, and all of the electrical connections to the DC electronic load and/or power supply.
The best-fit value was 0.125 ohm-cm$^{2}$.
(3) The diffusion layer thickness, which represents the critical mass transport parameter in the model, was chosen to be 5.85 $\mu \textrm{m}$ in order to best fit the limiting current density seen in the 70 psig case, and then this same value was used for all of the other pressures. This is why the model appears to more accurately match the cell's maximum power density in the 70 psig case than in the 12 psig case in Fig. \ref{fig:comparison-to-model}. Because the model utilizes a simple form of Henry's law to obtain the concentration of \ce{Cl2(aq)} in solution as a function of \ce{Cl2(g)} pressure, the limiting current density should be directly proportional to the absolute pressure of gas in the system. However, in Fig. \ref{fig:humid_complete_IV_PI} one can see that the limiting current density for the 12 psig case is approximately 0.65 A cm$^{-2}$, whereas the limiting current density for the 70 psig case is about 1.25 A cm$^{-2}$. The model predicts, based on the absolute pressure ratio, that these values should span a range of a factor of 3.2, whereas we find a range of less than a factor of two experimentally.
Clarifying the reasons for this behavior will require further research.

We show the overall fit of the model to the data in Figure \ref{fig:comparison-to-model}. Also shown for comparison are results from the``Base case'' and ``More Optimal case'' modeled by Rugolo \emph{et al.} \cite{Rugolo:2012kf} The latter is their prediction of the performance that may reasonably be expected with further research and development on a cell of this design. Table \ref{tab:parameter_comparison_table} directly compares parameter values of the ``Base case'' and the ``More Optimal case'' of Rugolo \emph{et al.}, and of the model fit to our experimental data. The experimental performance is well beyond that of the ``Base case'', and based on these results we anticipate that further R \&D could raise the power density at 90\% galvanic efficiency by a factor of three to four.

\begin{table*}
\small
  \caption{\ Parameters used in the ``Base case'', the ``More Optimal case'', and in fitting the model to the experimental data. Adjustable parameters are denoted by italics.}
  \label{tab:parameter_comparison_table}
  \begin{tabular*}{\textwidth}{@{\extracolsep{\fill}}r c c c c c c c c}
    \hline
    ~ & $i_0^{Cl}$ / mA cm$^{-2}$ & $i_0^H$ / mA cm$^{-2}$ & $\epsilon$ / $\mu \textrm{m}$ & L / $\mu \textrm{m}$ & P$_{\textrm{gauge}}$ / atm & R$_{\textrm{stack}}$A / ohm-cm$^{2}$ & Power at 90\% Efficiency / W cm$^{-2}$\\
    \hline
    \textbf{Base} & 10 & 250 & 3 & 150 & 1 & 0 & 0.1 \\
    \textbf{Fit} & \emph{175} & 250 & \emph{5.85} & 50 & 1.8-5.8 & \emph{0.125} & 0.285-0.358 \\
    \textbf{More Optimal} & 250 & 600 & 1 & 25 & 5 & 0 & 1.2 \\
    \hline
  \end{tabular*}
\end{table*}

\section{Conclusions}
We have developed a high-performance hydrogen-chlorine regenerative fuel cell. It incorporates a \ce{(Ru_{0.09}Co_{0.91})_3O_4} alloy as a chlorine redox electrocatalyst. We observed no significant activation losses, even with chlorine electrode precious metal loadings as low as 0.15 mg Ru cm$^{-2}$ . The peak galvanic power density exceeded 1 W cm$^{-2}$, which is twice that of previous work with much higher precious metal loadings on the \ce{Cl2} electrode. A power density of nearly 0.4 W cm$^{-2}$ was obtained at 90\% voltage efficiency, which is an important figure of merit when considering these devices for grid-scale electrical energy storage. The effects of \ce{Cl2} gas pressure, electrolyte acid concentration, and hydrogen electrode humidification were reported. We compare the results to the \ce{H2}-\ce{Cl2} fuel cell model of Rugolo \emph{et al.}, to which we added a series resistance term to better fit the experimental results seen here. The experimental performance is well beyond that of the ``Base case'' in the model, and the comparison to the model indicates the R\&D directions needed for another factor of 3-4 improvement in power density in order to reach the envisioned ``More Optimal'' case. Based on the high power densities and high efficiencies demonstrated here, we anticipate that a device such as this could be used in a flow battery configuration as a grid-scale electrical energy storage device. Further studies, including durability assessments and system-level integration, are necessary to determine its economic feasibility in this context. In the longer term, the device may become a component of a carbon sequestration scheme that mimics the natural chemical weathering process for \ce{CO2} removal from the atmosphere.

\section{Acknowledgments}

Several people contributed to this work. B.H. designed and built the cell on which measurements are reported, performed the measurements, and analyzed the results. J.R. designed an earlier generation cell upon which the final design was based, and helped design and build the measurement system. S.M. developed the alloy electrocatalyst composition and characterized its performance in half-cell measurements. M.J.A. directed the research. B.H. and M.J.A. wrote the manuscript. We thank Dr. Trent M. Molter for helpful discussions, and Josh Preston for help in designing, building, and troubleshooting issues with the measurement system. This research was partially supported by National Science Foundation grant NSF-IIP-0848366 through Sustainable Innovations, LLC. B.H. was supported by an NSF Graduate Research Fellowship.

\footnotesize{
\bibliography{FC_manuscript_refs} 

\begin{thebibliography}{10}

\bibitem{Mellentine:2011gw}
J.~A. Mellentine, W.~J. Culver, and R.~F. Savinell, ``{Simulation and
  optimization of a flow battery in an area regulation application},'' {\em
  Journal of Applied Electrochemistry}, vol.~41, pp.~1167--1174, June 2011.

\bibitem{Eyer:2010ti}
J.~Eyer and G.~Corey, ``{Energy storage for the Electricity Grid: Benefits and
  Market Potential Assessment Guide},'' tech. rep., Feb. 2010.

\bibitem{Rugolo:2012kj}
J.~Rugolo and M.~J. Aziz, ``{Electricity storage for intermittent renewable
  sources},'' {\em Energy {\&} Environmental Science}, vol.~5, p.~7151, 2012.

\bibitem{Gileadi:1977un}
E.~Gileadi, S.~Srinivasan, F.~Salzano, C.~Braun, A.~Beaufrere, S.~Gottesfeld,
  L.~Nuttall, and A.~Laconti, ``{An electrochemically regenerative
  hydrogen--chlorine energy storage system for electric utilities},'' {\em
  Journal of Power Sources}, vol.~2, no.~2, pp.~191--200, 1977.

\bibitem{Savinell:1988tn}
R.~Savinell and S.~Fritts, ``{Theoretical performance of a hydrogen-bromine
  rechargeable SPE fuel cell},'' {\em Journal of Power Sources}, vol.~22,
  no.~3-4, pp.~423--440, 1988.

\bibitem{Yeo:1980vr}
R.~Yeo and D.~Chin, ``{A Hydrogen‐Bromine Cell for Energy Storage
  Applications},'' {\em Journal of The Electrochemical Society}, vol.~127,
  p.~549, 1980.

\bibitem{Yeo:1980wr}
R.~Yeo, J.~McBreen, A.~Tseung, S.~Srinivasan, and J.~McElroy, ``{An
  electrochemically regenerative hydrogen-chlorine energy storage system:
  electrode kinetics and cell performance},'' {\em Journal of Applied
  Electrochemistry}, vol.~10, no.~3, pp.~393--404, 1980.

\bibitem{Thomassen:2006do}
M.~Thomassen, E.~Sandnes, B.~B{\o}rresen, and R.~Tunold, ``{Evaluation of
  concepts for hydrogen -- chlorine fuel cells},'' {\em Journal of Applied
  Electrochemistry}, vol.~36, pp.~813--819, July 2006.

\bibitem{Chin:1979uk}
D.~Chin, R.~Yeo, J.~McBreen, and S.~Srinivasan, ``{An Electrochemically
  Regenerative Hydrogen‐Chlorine Energy Storage System},'' {\em Journal of
  The Electrochemical Society}, vol.~126, p.~713, 1979.

\bibitem{Livshits:2006}
V.~Livshits, A.~Ulus, and E.~Peled, ``{High-power H2/Br2 fuel cell},'' {\em
  Electrochemistry communications}, vol.~8, pp.~1358--1362, 2006.

\bibitem{House:2009dm}
K.~Z. House, C.~H. House, D.~P. Schrag, and M.~J. Aziz, ``{Electrochemical
  acceleration of chemical weathering for carbon capture and sequestration},''
  {\em Energy Procedia}, vol.~1, pp.~4953--4960, Feb. 2009.

\bibitem{House:2007kw}
K.~Z. House, C.~H. House, D.~P. Schrag, and M.~J. Aziz, ``{Electrochemical
  Acceleration of Chemical Weathering as an Energetically Feasible Approach to
  Mitigating Anthropogenic Climate Change},'' {\em Environmental Science {\&}
  Technology}, vol.~41, pp.~8464--8470, Dec. 2007.

\bibitem{Thomassen:2005uc}
M.~Thomassen, B.~B{\o}rresen, G.~Hagen, and R.~Tunold, ``{Chlorine reduction on
  platinum and ruthenium: the effect of oxide coverage},'' {\em Electrochimica
  Acta}, vol.~50, no.~5, pp.~1157--1167, 2005.

\bibitem{Gasteiger:2004vw}
H.~Gasteiger, J.~Panels, and S.~Yan, ``{Dependence of PEM fuel cell performance
  on catalyst loading},'' {\em Journal of Power Sources}, vol.~127, no.~1-2,
  pp.~162--171, 2004.

\bibitem{Thomassen:2003wx}
M.~Thomassen, B.~B{\o}rresen, G.~Hagen, and R.~Tunold, ``{H2/Cl2 fuel cell for
  co-generation of electricity and HCl},'' {\em Journal of Applied
  Electrochemistry}, vol.~33, no.~1, pp.~9--13, 2003.

\bibitem{Neyerlin:2007ww}
K.~Neyerlin, W.~Gu, J.~Jorne, and H.~Gasteiger, ``{Study of the exchange
  current density for the hydrogen oxidation and evolution reactions},'' {\em
  Journal of The Electrochemical Society}, vol.~154, p.~B631, 2007.

\bibitem{Yeo:1979wv}
R.~Yeo and J.~McBreen, ``{Transport properties of Nafion membranes in
  electrochemically regenerative hydrogen/halogen cells},'' {\em Journal of The
  Electrochemical Society}, vol.~126, no.~10, pp.~1682--1687, 1979.

\bibitem{Nuttall:1977ta}
L.~Nuttall, J.~McElroy, S.~Srinivasan, and T.~Hart in {\em Proceeding of the
  Miami International Conference}, (Miami Beach, Fl), pp.~A79--34106 13--44,
  Dec. 1977.

\bibitem{Balko:1981tf}
E.~Balko, J.~McElroy, and A.~Laconti, ``{Halogen acid electrolysis in solid
  polymer electrolyte cells},'' {\em International Journal of Hydrogen Energy},
  vol.~6, no.~6, pp.~577--587, 1981.

\bibitem{Anderson:1994wn}
E.~Anderson, E.~Taylor, G.~Wilemski, and A.~Gelb, ``{A high performance
  hydrogen/chlorine fuel cell for space power applications},'' {\em Journal of
  Power Sources}, vol.~47, no.~3, pp.~321--328, 1994.

\bibitem{Bommaraju:2001tu}
T.~Bommaraju, C.~Chen, and V.~Birss, ``{Deactivation of Thermally Formed RuO2 +
  TiO2 Coatings During Chlorine Evolution: Mechanisms and Reactivation
  Measures},'' in {\em Modern Chlor-Alkali Technology} (J.~Moorhouse, ed.),
  pp.~57--81, London: Blackwell Science, Ltd., Jan. 2001.

\bibitem{Beer:1980vf}
H.~Beer, ``{The invention and industrial development of metal anodes},'' {\em
  Journal of The Electrochemical Society}, vol.~127, no.~8, pp.~303C--307C,
  1980.

\bibitem{Trasatti:1984wq}
S.~Trasatti, ``{Electrocatalysis in the anodic evolution of oxygen and
  chlorine},'' {\em Electrochimica Acta}, vol.~29, no.~11, pp.~1503--1512,
  1984.

\bibitem{Ardizzone:1982up}
S.~Ardizzone, A.~Carugati, G.~Lodi, and S.~Trasatti, ``{Surface Structure of
  Ruthenium Dioxide Electrodes and Kinetics of Chlorine Evolution},'' {\em
  Journal of The Electrochemical Society}, vol.~129, p.~1689, 1982.

\bibitem{Mondal:2011hd}
S.~K. Mondal, J.~Rugolo, and M.~J. Aziz, ``{Alloy Oxide Electrocatalysts for
  Regenerative Hydrogen-Halogen Fuel Cell},'' in {\em MRS Proceedings} (T.~He,
  K.~E. Swider-Lyons, B.~Park, and P.~Kohl, eds.), Jan. 2011.

\bibitem{Cullity:1956tk}
B.~Cullity, ``{The Determination of Crystal Structure},'' in {\em Elements of
  X-Ray Diffraction}, pp.~297--323, Addison-Wesley, Jan. 1956.

\bibitem{Beckmann:2001uo}
R.~Beckmann and B.~L{\"u}ke, ``{Know-how and Technology - Improving the Return
  on Investment for Conversions, Expansions, and New Chlorine Plants},'' in
  {\em Modern Chlor-Alkali Technology} (J.~Moorhouse, ed.), pp.~196--212,
  London: Blackwell Science, Ltd., Jan. 2001.

\bibitem{Rugolo:2012kf}
J.~Rugolo, B.~Huskinson, and M.~J. Aziz, ``{Model of Performance of a
  Regenerative Hydrogen Chlorine Fuel Cell for Grid-Scale Electrical Energy
  Storage},'' {\em Journal of The Electrochemical Society}, vol.~159, no.~2,
  p.~B133, 2012.

\end{thebibliography}
\bibliographystyle{ieeetr} 
}

\end{document}